\documentstyle[aps,psfig]{revtex}
\tighten
\draft

\newcommand{\bi}[1]{\bibitem{#1}}
\newcommand{\fr}[2]{\frac{#1}{#2}}

\newcommand{\gam}{\gamma}

\newcommand{\la}{\lambda}

\newcommand{\be}{\begin{equation}}
\newcommand{\ee}{\end{equation}}
\newcommand{\bea}{\begin{eqnarray}}
\newcommand{\eea}{\end{eqnarray}}

\def\ga{\mathrel{\raise.3ex\hbox{$>$\kern-.75em\lower1ex\hbox{$\sim$}}}}
\def\la{\mathrel{\raise.3ex\hbox{$<$\kern-.75em\lower1ex\hbox{$\sim$}}}}
\newcommand{\Mpl}{$M_{\rm {\small Pl}}$}

\newcommand{\ie}{{\it i.e.,}\ }
\newcommand{\eg}{{\it e.g.,}\ }
\newcommand{\reef}[1]{(\ref{#1})}
\newcommand{\cL}{{\cal L}}
\newcommand{\ssc}{\scriptscriptstyle}
\newcommand{\LL}[1]{\cL_{\ssc #1}}
\newcommand{\prt}{\partial}
\newcommand{\Phib}{\bar{\Phi}}
\newcommand{\Psib}{\bar{\Psi}}
\newcommand{\veps}{\varepsilon}
\newcommand{\eps}{\epsilon}
\newcommand{\mpl}{M_{\rm {\small Pl}}}
\newcommand{\sla}[1]{\hbox{{$#1$}\llap{$/$}}}
\newcommand{\sixth}{{{\scriptstyle 1}\over{\scriptstyle 6}}}

\begin{document}

\twocolumn[\hsize\textwidth\columnwidth\hsize\csname
@twocolumnfalse\endcsname

\title{Ultraviolet modifications of dispersion relations in effective field theory}
\author {Robert C.~Myers$^{1,2}$
 and
Maxim Pospelov$^{3,4}$}
\address{$^{1}$ Perimeter Institute for Theoretical Physics, 35 King
Street North, Waterloo, Ontario N2J 2W9, Canada}
\address{$^{2}$ Department of Physics, McGill University,
Montr\'eal, Qu\'ebec H3A 2T8, Canada}
\address{$^{3}$ Department of Physics and Astronomy,
University of Victoria, Victoria, BC, V8P 1A1, Canada}
\address{$^{4}$ Centre for Theoretical Physics, University of Sussex, Brighton
BN1 9QJ,~~UK}

\maketitle

\begin{abstract}

The existence of a fundamental ultraviolet scale, such as the
Planck scale, may lead to modifications of the dispersion
relations for particles at high energies, in some scenarios of
quantum gravity. We apply effective field theory to this problem
and identify dimension 5 operators that do not mix with dimensions
3 and 4 and lead to cubic modifications of dispersion relations
for scalars, fermions, and vector particles. Further we show that,
for electrons, photons and light quarks, clock comparison
experiments bound these operators at $10^{-5}$/\Mpl.

\end{abstract}

\vskip1pc]

\section{Introduction}

The Planck mass, \Mpl, a dimensional parameter determining the
strength of gravitational interaction, remains a source of
conceptual problems for quantum field theories. When the momentum
transfer in a process of two particle collision is comparable to
the Planck mass, the graviton exchange becomes strong, signifying
a breakdown of the perturbative field theory description. Even
without having a fully consistent fundamental theory at hand, one
can hypothesize several broad categories of the low-energy effects
induced by \Mpl. The first group of such theories has only minor
modifications due to the existence of new physics at \Mpl. By
minor modifications we understand that all ``sacred'' symmetries
(Lorentz symmetry, CPT, spin-statistics, etc.) of the field theory
remain unbroken at low energies. String theory in the critical
dimension and in the simplest background reduces to field theory
and gravity at the scales lower than $M_{s}$, and provides a
perfect example of this category. In this case the chances to
probe 1/\Mpl\ suppressed effects are very remote, as one has to
have a unrealistically large momentum transfers. Consequently, the
propagation of a free particle with large energy/momentum is
immune to the effects of new physics, as all corrections to the
dispersion relation could be cast in the form of
$(p^2)(p^{2n}/M_{\rm Pl}^{2n})= (m^{2n+2}/M_{\rm Pl}^{2n})$ where
$m$ is the mass of the particle and $p$ is the four-momentum.

In the second class of scenarios, that include loop quantum gravity,
the \Mpl\ effects have much more ``vivid'' properties.
In this approach, the
discrete nature of space at short distances is expected to induce
violations of Lorentz invariance and CPT. Such violations are also
often discussed in a broader context using field theoretical
language \cite{Kost}. Here one might assume a perfectly Lorentz
symmetric, CPT conserving action at $1/M_{\rm Pl}^{0}$ order, and
account for the existence of the Lorentz breaking terms via a set
of higher dimension operators. This would lead to modifications of
the dispersion relation for a free particle as terms of the form
$E^{n+2}/\mpl^{n}$ can appear. Such effects can be searched for
with astrophysical tools \cite{mass,Jacob,wow,rot} or with high
precision low-energy experiments \cite{terra,Vuc}.

In this Letter, we consider cubic modifications of dispersion
relations, which would appear at the leading order 1/\Mpl~ and
which have received considerable attention in the literature
recently \cite{mass,Jacob,wow,rot}. In the language of the
effective field theory, such modifications can be described by the
dimension 5 operators. Although dimension 3,4 operators were
extensively studied \cite{Kost}, dimension 5 operators remain
unclassified and poorly explored. Lorentz breaking can be achieved
by the introduction of an ``external'' background with open
Lorentz indices. While more complicated choices are conceivable,
we consider the case where a preferred frame is defined by a fixed
time-like four-vector $n^a$. Within this framework, we study cubic
modifications of the dispersion relation for scalars, vectors and
fermions. Specializing the operators to Standard Model particles,
we show that the cubic modification is not possible for the Higgs
particles, must have an opposite sign for opposite chiralities of
photons, and is independent for different chiralities of fermions.
Note that the commonly studied dispersion relations
\cite{mass,Jacob,wow} do not conform with those derived here.

Further we consider certain experimental bounds on these
operators. It is important to realize that the introduction of
{\em any} preferred frame leads to spatial anisotropy. The
four-vector $n^a$ that characterizes the preferred frame may have
a dominant component $n^0$ in some cosmic frame, but the motion of
our galaxy, Solar system and Earth will create spatial components
$n_i \sim 10^{-3}$ for a terrestrial observer \cite{nuCPT}. Hence
clock comparison experiments \cite{clocks} or experiments with
spin-polarized matter \cite{Heckel} searching for the  spatial
anisotropy can impose stringent bounds on violations of Lorentz
symmetry in this context. Recently direction-sensitive clock
comparison results were used to constrain the dispersion relation
for nucleons \cite{Vuc}. Here with a direct application of the
spin precession bounds, we derive $O(1)$ limits for leptons,
$O(10^{-5})$ for photons and $O(10^{-8})$ for quarks in units of
1/\Mpl. The results in the lepton sector can be improved to the
level of $10^{-5}\mpl^{-1}$ with a renormalization group analysis.
Comparing to recent calculations \cite{loop1,loop2,loop3}, we find
that these bounds already present an interesting challenge for
loop quantum gravity.

\section{Cubic modifications of dispersion relations by dimension
5 operators}

In the framework of low energy effective field theory, the
modified dispersion relations should be derived from some
appropriate modification of the kinetic terms in the Lagrangian.
In this paper we shall assume that short-distance physics does not
generate dimension three and four terms directly. Even though this
assumption has serious theoretical problems (\eg one wants a
symmetry to forbid such operators), it is quite safe from the
practical side, because even if these terms exist they are already
severely constrained by experiment (see, \eg \cite{Kost}).

At the next level are dimension five operators which would lead to
$O(E^3)$ modifications of the dispersion relations. We adopt the
simplest approach where Lorentz symmetry is broken a background
four vector $n^a$ (with $n\cdot n=1$). We construct operators
satisfying six generic criteria:\\
$~~~$1. Quadratic in the {\em same} field\\
$~~~$2. One more derivative than the usual kinetic term\\
$~~~$3. Gauge invariant\\
$~~~$4. Lorentz invariant, except for the appearance of $n^a$\\
$~~~$5. Not reducible to lower dimension operators by \\\phantom{$~~~~~$}
 the equations of motion\\
$~~~$6. Not reducible to a total derivative

Conditions 2 and 5 ensure that these operators lead to $O(E^3)$
modifications of the dispersion relations, rather than $O(E^2m)$
or $O(Em^2)$, where $m$ is the mass of the particle. Our working
assumption will also be that these operators are naturally
suppressed by a factor of 1/\Mpl, and that $m,E \ll M_{\rm
Pl}$. This scaling ensures that all operators of dimension five
can be regarded as small perturbations. We shall consider cases of
a scalar, fermion and vector particle.

\noindent{\bf Scalar:} We work with a complex scalar field with
the standard dimension four Lagrangian: $ \LL0 =
|\prt\Phi|^2-m^2|\Phi|^2.$ Hence the leading order equation of
motion yields $(\Box+m^2)\Phi=0$, or in momentum space with
$\Phi\sim\exp(-ik\cdot x)$, $(-k^2+m^2)\Phi(k)=0$.

It is relatively easy to see that there is only one possible term
which appears at the dimension five level
\be \label{nexts}
{\cal L}_s=
i{\kappa\over\mpl}\Phib(n\cdot\prt)^3\Phi
\ee
which is odd under $CPT$ and charge conjugation. The
corresponding equation of motion becomes
\be\label{neweoms}
(\Box+m^2)\Phi=i{\kappa\over\mpl}(n\cdot\prt)^3\Phi.
\ee
Another possible operator, $\Phib n\cdot\prt\Box\Phi$, is
reducible to $m^2 \Phib n\cdot\prt\Phi$ and does not effect the
dispersion relation in a significant way. In momentum space (with
$n\cdot\prt\sim -iE$), and in the Lorentz frame where $n^a =
(1,0,0,0)$, we have the dispersion relation
\be E^2
\simeq|\vec{p}|^2+m^2 +{\kappa\over\mpl}|\vec{p}|^3
\label{newdisps} \ee
where we have used $E\simeq|\vec{p}|$ for high energies.

Note that for a real scalar (like the Standard Model Higgs), the
operator ${\cal L}_s$ is the total derivative and does not produce
a cubic modification of the dispersion relation. For the case of
complex scalar, the introduction of $\LL{s}$ respects the phase
invariance of $\LL0$. The dimension five operator  breaking this
symmetry, \eg $\Phi(n\cdot\prt)^3\Phi$ vanishes again as a total
derivative and is forbidden if the phase symmetry is gauged.

\noindent{\bf Vector:} We work with a U(1) gauge field for which
the leading kinetic term is $\LL0 = -
F^2/4$. (Extending the following to a nonabelian vector is
straightforward.) Hence the leading order equations of motion are
just the
Maxwell equations: $\prt_a F^{ab}=0$. After gauge fixing
$\prt\cdot A=0$, this yields $\Box A_a=0$, or in momentum space
$k^2\,A_a(k)=0$. Again, we wish to
modify the dispersion relation at $O(E^3)$ and so the new terms
should satisfy the constraints listed above. Keeping in mind the
leading order Maxwell equations in vacuum and the Bianchi
identities $\prt_{[a}F_{bc]}=0$, one finds that there is only one
term with the desired form and which produces a nontrivial
modification of the dispersion relations
\be \label{nextv} {\cal L}_\gamma=
{\xi\over\mpl} n^a F_{ad} n\cdot\prt(n_b\tilde F^{bd}), \ee
where $\tilde F^{ab}= \fr{1}{2} \veps^{abcd}F_{cd}$. Note that
this operator is odd under $CPT$ and even under charge
conjugation. The equation of motion becomes
\be\label{neweomv}
\Box A_a=-{\xi\over\mpl}\veps_{abcd}n^b (n\cdot\prt)^2F^{cd}
\ee
where we used Bianchi identity and $\prt\cdot A=0$.
To identify the effect of the new term on the dispersion
relation, we go to momentum space and select photons moving along
the $z$ axis with $k^a=(E,0,0,p)$. Then for transverse
polarizations along the $x$ and $y$ axes,
\be \label{newdispv}
\left(E^2-p^2\pm{2\xi\over\mpl}p^3\right)(\eps_x\pm i\eps_y)
= 0 \ee
where we have used $E\simeq p$ to leading order and chosen the
``rest frame'' for $n^a$ as before. Hence the sign of the cubic
term is determined by the chirality (or circular polarization) of
the photons. As in the case of optical activity, this leads to the
rotation of the plane of polarization for linearly polarized
photons, which is used to bound $\xi$ in \cite{rot}. Given that
$\LL\gamma$ is unique, the common approach \cite{mass,Jacob,wow}
of postulating a cubic dispersion relation which is chirality
independent is incompatible with effective field theory.

\noindent{\bf Spinor:} We work with a Dirac spinor for which the
leading kinetic term is: $\LL0 = \Psib(i\sla{\prt}-m)\Psi.$
Again, we
wish to modify the dispersion relation at $O(E^3)$ and so consider
new terms satisfying the constraints listed above. Keeping in mind
the leading order equations, one finds that there are only two
terms with the desired form
\be \label{nextf} {\cal L}_f={1\over\mpl}\Psib\left( \eta_1\sla{n}
+\eta_2\sla{n}\gam_5
\right)(n\cdot\prt)^2\Psi \ee
Both operators break CPT, with $\eta_1$ being charge conjugation
odd and $\eta_2$ charge conjugation even. Now the equation of
motion takes the form
\be\label{neweomf} (i\sla{\prt}-m)\Psi= -{1\over\mpl}\left(
\eta_1\sla{n} 
+\eta_2\sla{n}\gam_5 
\right)(n\cdot\prt)^2\Psi\ee
To produce an expression that is readily identified with a
dispersion relation, one operates on both sides with
$(i\sla{\prt}+m)$ to produce
\be\label{neweomf2} (\Box+m^2)\Psi= {2i\over\mpl}\left(\eta_1
+\eta_2\gam_5\right)(n\cdot\prt)^3\Psi\ee
where we have again dropped terms of order $m/\mpl$.
Hence the modified dispersion relation becomes
\be \label{newdispf}
\left(E^2-|\vec{p}|^2-m^2-{2|\vec{p}|^3\over\mpl}(\eta_1
+\eta_2\gam_5)\right)\Psi =0 \ee
with $E\simeq|\vec{p}|$ for high energies. At high energies (\ie
$E^2\gg m^2$), we can choose spinors as eigenspinors of the
chirality operator and redefine coupling constants as $\eta_{R,L}
= \eta_1 \pm \eta_2$. In the extension of the Standard Model, the
chiral choice for $\eta$ couplings is prescribed by gauge
invariance. Previous studies \cite{Jacob,wow} consider only
chirality independent dispersion relations for fermions and so
implicitly fix $\eta_2=0$.

Above we have identified interesting operators which modify the
dispersion relations at cubic order. One could also consider
frame-dependent modifications of the interaction terms between,
\eg photons and electrons. However, these would not effect the
threshold tests which are usually studied, \eg \cite{Jacob}. On
the other hand, they may play a role in indirect tests as those
considered below.

As a further technical aside, one can see a very serious potential
problem with divergences at the loop level. For example, it is
easy to show that when inserted in a self-energy loop, the
$\eta_2$ operator generates a dimension 3 operator
$\Lambda^2_{UV}\mpl^{-1} \bar \psi \sla{n}\gamma_5\psi $ with
$\Lambda_{UV}$, the ultra-violet cutoff on the momentum integral.
The latter cannot be lower than the electroweak scale or SUSY
breaking scale. However, even assuming such a low cutoff, the
bounds of \cite{Kost} on these dimension three operators restrict
all the couplings, $\eta_i, \xi, \kappa$ to be smaller than
$10^{-10}$. However, we may evade this bound as follows: The
external tensor $n^an^bn^c$ appearing in all of the operators
above may be separated into a vector piece and a traceless
symmetric tensor $C^{abc}= n^an^bn^c -\sixth$($n^ag^{bc}+$cyclic).
The quadratic divergences above are associated with the vector
part only, while the operators proportional to $C^{abc}$ can
generate only logarithmic divergences which should be interpreted
as the renormalization group evolution of these operators. Thus,
to be consistent with the assumption about the absence of large
dimension 3 or 4 operators, we make the substitution $n^an^bn^c
\rightarrow C^{abc}$ in the following. Note that this change does
not affect the dispersion relations in the regime $E\gg m$.

\section{Low-energy limits on dimension five operators}

Modifications of dispersion relations for stable particles such as
electrons, light quarks, and photons could be searched for using
the astrophysical probes \cite{mass,Jacob,wow,rot}. Here we show
that terrestrial limits imposed by the clock comparison
experiments are equally and sometimes
more sensitive to dimension 5 operators. Indirect
limits exploit the idea that the choice of $n^a=(1,0,0,0)$ defines
a preferred frame that does not coincide with the laboratory frame
on the Earth \cite{nuCPT}. Assuming that the rest frame for $n^a$
is related somehow to either the cosmic or galactic frame, the
typical size of the spatial component of $n$ is $10^{-3}$.

Limits on the operators involving electrons and electron neutrinos
are especially easy to derive. We use the fact that best tests of
directional sensitivity in the precession of electrons limit the
size of interaction between the external direction and the
electron spin at the level of $10^{-28}$ GeV \cite{Heckel}. This
immediately translates into the following limit on the
coefficients $\eta_L$ and $\eta_R$ that parametrize the effective
interaction of the form \reef{nextf} for left-handed leptons and
right-handed electrons:
\be |\eta_L^e-\eta_R^e|\la
\fr{10^{-28}~{\rm GeV}\mpl}{m_e^2|n_i|}\simeq 4, \label{leptl}
\ee
where we use $\mpl \equiv 10^{19}$ GeV. The square of the electron
mass originates from $(n^0\partial_t)^2$ acting on the electron
wave function. The orthogonal combination, $\eta_L^e+\eta_R^e$, is
almost not constrained, as it does not contribute into the
electron spin Hamiltonian. Similarly, we do not constrain
operators involving second and third generation leptons. Note that
our limit is comparable to the existing constraints \cite{Jacob}.
However the latter analysis assumed $\eta_2=0$ and so our result
is a complementary constraint.

The absence of a preferred direction is checked with even greater
precision using nuclear spin, which translates into more
stringent limits on new operators for the light quarks. The photon
operator (\ref{nextv}) will also contribute because of the
electromagnetic interactions inside the nucleon. To use the best
experimental limits of $10^{-31}$ GeV on the coupling of $n_i$ to
neutron spins \cite{clocks}, we must relate the photon and quark
operators with nucleon spin. Introducing dimension five operators
for left-handed doublet $\psi_Q$ and right-handed singlets $\psi_u$ and
$\psi_d$, members of the first generation of quarks
\bea
{\cal L}_q &=& \frac{C^{abc}}{\mpl}
\sum_{i=Q,u,d}\eta_i \bar \psi_i \gamma_a \prt_b\prt_c \psi_i,
\label{Qud}
\eea
at the nucleon level we have
\begin{eqnarray}
\label{matel}
\eta_{1,N} = a_{u}(\eta_u +\eta_Q)+a_{d}(\eta_d +\eta_Q)\\
\eta_{2,N} =b_{u}(\eta_u -\eta_Q)+b_{d}(\eta_d -\eta_Q) + b_{\gamma}\xi,
\nonumber
\eea
where $\eta_{1(2),N}$ are the $\eta_1$ and $\eta_2$ couplings for
nucleons defined in (\ref{nextf}).
Note that $\xi$ enters only in the $\eta_2$ coupling for nucleons
because both are even under the charge conjugation.
In (\ref{matel}) $a_{u,d}$ and $b_{u,d}$ are the
matrix elements that could be obtained as the moments of the
experimentally measured structure functions:
$a_d \sim 0.4  $, $a_u \sim 0.1 $, $b_d\sim 0.1 $, $b_u\sim -0.05$
for the neutron and charge inverted values for the proton. To
relate the photon operator with nucleon, we use the simplest
vector dominance model and obtain at one-loop level $b_\gamma \sim
0.13\alpha/(4 \pi)$ for neutron and $b_\gamma \sim 0.24 \alpha/(4
\pi)$ for proton. This results in the 
limit:
\begin{eqnarray}
|(\eta_d - \eta_Q) - 0.5 (\eta_u - \eta_Q) + 10^{-3}\xi| \la
10^{-8} \label{etaquark}
\end{eqnarray}
Barring accidental cancellations, one can place separate limits on
$\eta_{u,d} - \eta_Q$ at $10^{-8}$ and on $\xi$ at $10^{-5}$
level. The orthogonal combinations $\eta_{u,d} + \eta_Q$ are less
constrained because they enter only in the quadrupole coupling
between the nuclear spin and external direction, and thus are
suppressed by an additional factor of $|n_i|\sim 10^{-3}$
\cite{Vuc}.

So far we have neglected the fact that the low energy values for
the couplings $\eta_i$ and $\xi$ taken at the normalization scale
of 1 GeV do not coincide with the high-energy values for the same
couplings generated at $\mpl$. With a simple one-loop analysis of
the renormalization group equations, we find that several bounds
can be strengthened. Leaving the details for elsewhere, our
results are: $|\eta_{Q,u,d}|,|\xi| \la 10^{-6}$ and
$|\eta_{L,R}^e| \la 10^{-5}$ with the same normalization on
$1/\mpl$. The constraints may also be improved by assuming a GUT
scenario.

In summary, we have shown that effective field theory provides a
framework where one can derive stringent bounds on Planck scale
interactions from terrestrial experiments. The resulting limits
enhance and generalize the bounds obtained previously in \cite{Vuc}. They
are generally far more sensitive \cite{Vuc} than those previously derived by
considering astrophysical phenomena \cite{Jacob} where the typical
sensitivity is $O(1)$ in units of $\mpl^{-1}$.
Actually the birefringence induced by $\LL\gamma$ can be used to bound
$|\xi|\la10^{-4}$ with astronomical observations \cite{rot}, but
our bound improves on this result by roughly an order of
magnitude. Recently \cite{wow} infers a striking bound
$\eta_{L}^e+\eta_{R}^e\ga-10^{-9}$ (assuming
$\eta_{L}^e-\eta_{R}^e=0$) from the observation of synchrotron
radiation from astrophysical sources. Here our results provide an
upper bound for positive couplings.

Following \cite{Vuc}, we compare our results to semi-classical calculations
which have appeared using loop quantum gravity
\cite{loop1,loop2,loop3}. Examining the interactions induced for a
Dirac fermion \cite{loop2} shows that $\eta_1=0$ and that while
$\eta_2$ is nonvanishing, it is suppressed an additional small
parameter related to the coherence scale of the gravitational wave
function. Hence these calculations seem to be in accord with the
stringent bounds derived for the fermion operators. In contrast,
the results of \cite{loop1} suggest that $\xi$ should be $O(1)$,
which stands in stark contradiction with the bound derived here.
Hence our bounds seem to be in conflict with the present
calculations \cite{loop1,loop2,loop3}. However, the latter must be
viewed as somewhat heuristic. So the immediate challenge for loop
quantum gravity is to provide rigorous predictions that may be
compared to experiment. Despite the many unresolved theoretical
issues, it is truly remarkable that precision experiments can
already confront quantum gravity calculations with concrete
observational bounds.

We thank Giovanni Amelino-Camelia, Cliff Burgess, Ted Jacobson,
Joe Lykken, Guy Moore, Subir Sarkar, Lee Smolin and especially
Seth Major for useful conversations. This research is supported in part by
NSERC of Canada,  Fonds FCAR du Qu\'ebec and PPARC UK.


\begin{thebibliography}{99}

\bi{Kost} D.~Colladay and V.A.~Kostelecky, Phys.\ Rev.\ {\bf
D55} (1997) 6760; 
S.R.~Coleman and S.L.~Glashow, Phys.\ Rev.\ {\bf D59} (1999)
116008; 
S.M.~Carroll, G.B.~Field and R.~Jackiw, Phys.\ Rev.\ {\bf D41}
(1990) 1231.

\bi{mass} G.~Amelino-Camelia {\it et al.}, Nature {\bf 393} (1998)
763;
G.~Amelino-Camelia and T.~Piran, Phys.\ Rev.\ {\bf D64} (2001)
036005; 
G.~Amelino-Camelia, Phys.\ Lett.\ {\bf B528} (2002) 181;
S.~Sarkar, Mod.\ Phys.\ Lett.\ {\bf A17} (2002) 1025.

\bibitem{Jacob}
T.~Jacobson, S.~Liberati and D.~Mattingly, Phys.\ Rev.\ {\bf D66}
(2002) 081302;  
 hep-ph/0209264;
T.J.~Konopka and S.A.~Major,
New J.\ Phys.\  {\bf 4} (2002) 57.

\bi{wow} T.~Jacobson, S.~Liberati and D.~Mattingly,
astro-ph/0212190.

\bi{rot} R.J.~Gleiser and C.N.~Kozameh, Phys.\ Rev.\ {\bf D64}
(2001) 083007.

\bi{terra} G.~Amelino-Camelia, Nature {\bf 398} (1999) 216;
 I.~Mocioiu, M.~Pospelov
and R.~Roiban, Phys.\ Lett.\ {\bf B489} (2000) 390;
G.Z. Adunas, E.~Rodriguez-Milla and D.V.~Ahluwalia, Phys.\ Lett.\
{\bf B485} (2000) 215; 
R.~Bluhm,  hep-ph/0111323.


\bi{Vuc} D.~Sudarsky, L.~Urrutia and H.~Vucetich, Phys.\ Rev.\
Lett.\  {\bf 89} (2002) 231301.


\bibitem{nuCPT}
I.Mocioiu and M.Pospelov, Phys.\ Lett.\ {\bf B534} (2002) 114.

\bibitem{clocks}
T.E. Chupp {\it et al.}, Phys. Rev. Lett. {\bf 72}, 2363 (1994);
R.E. Stoner {\it et al.}, Phys. Rev. Lett. {\bf 77}, 3971 (1996);
D. Bear {\it et al.}, Phys. Rev. {\bf A57}, 5006 (1995); C.J.
Berglund {\it et al.}, Phys. Rev. Lett. {\bf 75}, 1879 (1995);
D.~Bear, {\it et al.} Phys.\ Rev.\ Lett.\  {\bf 85} (2000) 5038
[Errat.\  {\bf 89} (2002) 209902].

\bibitem{Heckel}
E.G. Adelberger {\it et al.}, in {\it Physics Beyond the Standard
Model, 1998, Santa Fe} (World Scientific, Singapore, 1999); M.G.
Harris, PhD Thesis, University of Washington, 1998; B.R. Heckel
{\it et al.}, in {\it Proceedings of the International Conference
on Orbis Scientiae, 1999, Coral Gables} (Kluwer, New York, 2000).

\bi{loop1} R.~Gambini and J.~Pullin, Phys.\ Rev.\ {\bf D59} (1999)
124021;
J.~Alfaro, H.A.~Morales-Tecotl and L.F.~Urrutia, Phys.\ Rev.\ {\bf
D65} (2002) 103509.

\bi{loop2} J.~Alfaro, H.A.~Morales-Tecotl and L.F.~Urrutia,
 hep-th/0208192.

\bi{loop3} H.~Sahlmann and T.~Thiemann, gr-qc/0207030;
gr-qc/0207031.




\end{thebibliography}
\end{document}